\documentclass[conference]{IEEEtran}
\ifCLASSINFOpdf
\else
\fi

\usepackage{amssymb}
\usepackage{latexsym}
\usepackage{graphicx}
\usepackage{color}
\usepackage{verbatim}
\usepackage{multirow}
\usepackage{amsmath}

\usepackage{amsfonts}

\newcommand{\F}{\ensuremath{\mathbb F}}

\newcommand{\C}{\ensuremath{\mathbb C}}
\newcommand{\R}{\ensuremath{\mathbb R}}

\newcommand{\abu}{{\bf a}}

\newcommand{\sbu}{{\bf s}}

\newcommand{\rbu}{{\bf r}}

\newcommand{\xbu}{{\bf x}}
\newcommand{\ybu}{{\bf y}}

\newcommand{\zbu}{{\bf z}}

\newcommand{\Abu}{{\bf A}}

\newcommand{\Gbu}{{\bf G}}

\newcommand{\qed}{\hfill \ensuremath{\Box}}

\newtheorem{prop}{Proposition}
\newtheorem{df}{Definition}
\newtheorem{thr}{Theorem}

\newtheorem{rem}{Remark}

\newtheorem{const}{Construction}

\begin{document}
%
\title{Deterministic Compressed Sensing Matrices from Multiplicative Character Sequences}


\author{\IEEEauthorblockN{Nam Yul Yu} \\
\IEEEauthorblockA{Department of Electrical Engineering, Lakehead University \\
Thunder Bay, ON, CANADA \\
Email: nam.yu@lakeheadu.ca \\
}
}


%


\maketitle

\begin{abstract}
Compressed sensing is a novel technique where one can
recover sparse signals from the
undersampled measurements.
In this paper, a $K \times N$ measurement matrix for compressed sensing
is deterministically constructed
via multiplicative character sequences.
Precisely, a constant multiple of a cyclic shift of an $M$-ary power residue or Sidelnikov sequence is arranged
as a column vector of the matrix, through modulating a primitive $M$-th root of unity.
The Weil bound is then used to show that the matrix has asymptotically optimal coherence for large $K$ and $M$,
and to present a sufficient condition on the sparsity level for unique sparse solution.
Also, the restricted isometry property (RIP) is statistically studied for the deterministic matrix.
Numerical results
show that the deterministic compressed sensing matrix guarantees reliable matching pursuit recovery performance
for both noiseless and noisy measurements.
\end{abstract}

\begin{keywords}
Compressed sensing,
multiplicative characters,
power residue sequences,
restricted isometry property,
Sidelnikov sequences,
Weil bound.
\end{keywords}

%

\section{Introduction}
Compressed sensing (or compressive sampling) is a novel and emerging technology
with a variety of applications in imaging, data compression, and communications.
In compressed sensing, one can recover sparse signals of high dimension
from few measurements that were believed to be incomplete.
Mathematically, measuring an $N$-dimensional signal $\xbu \in \R^N$
with a $K \times N$ measurement matrix $\Abu$ produces
a $K$-dimensional vector $\ybu = \Abu \xbu $ in compressed sensing, where $K<N$.
In recovering $\xbu$, it seems impossible to solve $K$ linear equations
with $N$ indeterminates by basic linear algebra.
However, imposing an additional requirement that $\xbu$ is $s$-sparse
or the number of nonzero entries in $\xbu$ is at most $s$,
one can recover $\xbu$ exactly with high probability
by solving the $l_1$-minimization problem,
which is computationally tractable.

Many research activities have been triggered on theory and practice of compressed sensing
since Donoho~\cite{Donoho:CS}, and Candes, Romberg, and Tao~\cite{CanRomTao:robust}\cite{CanTao:univ}
published their marvelous theoretical works.
The efforts revealed that
a measurement matrix $\Abu$ plays a crucial role in recovery of
$s$-sparse signals. 
In particular, Candes and Tao~\cite{CanTao:univ} presented the \emph{restricted isometry property (RIP)},
a sufficient condition for the matrix to guarantee sparse recovery. 
A \emph{random} matrix has been widely studied for satisfying the RIP,
where the entries are generated by a probability distribution such as the Gaussian or Bernoulli process,
or from randomly chosen partial Fourier ensembles.

Although a random matrix has many theoretical benefits~\cite{Rauhut:struct}, 
it has the drawbacks~\cite{Cald:large}
of high complexity, large storage, and low efficiency 
in its practical implementation.
As an alternative, 
we may consider 
a \emph{deterministic} matrix, where well known codes and sequences have been employed for the construction, 
e.g., chirp sequences~\cite{App:chirp}, Kerdock and Delsarte-Goethals codes~\cite{Cald:sub},
second order Reed-Muller codes~\cite{How:fast}, and dual BCH codes~\cite{Ailon:BCH}.
Other techniques for deterministic construction, based on finite fields, representation theory, additive characters,
and cyclic difference sets,
can be found in~\cite{DeVore:det}$-$\cite{Yu:CDS}.
Although it is difficult to check the RIP 
and the theoretical recovery bounds are worse than that of a random matrix~\cite{Rauhut:struct},
deterministic matrices guarantee reliable recovery performance in a statistical sense,
allowing fast processing and low complexity.

To enjoy the benefits of deterministic construction,
this paper presents how to construct a $K \times N$ measurement matrix for compressed sensing
via \emph{multiplicative character sequences}.
Precisely, we construct the matrix
where a constant multiple of a cyclic shift of an $M$-ary power residue or Sidelnikov
sequence of length $K$ is arranged as a column vector, through modulating a primitive $M$-th root of unity.
The Weil bound is then used to show that the matrix has asymptotically optimal coherence for large $K$ and $M$, and
to present the theoretical bound on the sparsity level 
for unique sparse solution.
The RIP of the matrix is also analyzed through the eigenvalue statistics of the Gram matrices as in~\cite{Cald:large} and~\cite{App:chirp}.
Through numerical results,
we observe that the matching pursuit recovery for our deterministic matrices
provides stable and reliable performance in recovering sparse signals with or without measurement noise.

The rest of this paper is organized as follows.
In Section II, we describe the background for understanding this work.
Section III presents the main contribution of this paper
by constructing compressed sensing matrices from multiplicative character sequences,
and studying the properties on sparse recovery.
In Section IV, numerical results of the recovery performance are given
for noiseless and noisy measurements.
Concluding remarks will be given in Section V.


\section{Preliminaries}
The following notations will be used throughout this paper.
\begin{enumerate}
\item[$-$] $\omega_M = e^{j \frac{ 2 \pi}{M}}$ is a primitive $M$-th root of unity, where $j = \sqrt{-1}$.
\item[$-$] $\F_q={\rm GF}(q)$ is the finite field with $q$ elements and 
$\F_q^*= \F_q \setminus \{0 \}$ denotes the multiplicative group of $\F_q$.
\item[$-$] $\F_q [x]$ is the polynomial ring over $\F_q$, where each coefficient of $f(x) \in \F_q[x]$ is an element of $\F_q$.
\item[$-$] For $x$ in $\F_q$,
a \emph{logarithm} over $\F_q$ is defined by
\[
\log_\alpha x = \left \{ \begin{array}{ll} t, & \quad \mbox{if } x = \alpha^t, \ 0 \leq t \leq q-2, \\
0, & \quad \mbox{if } x = 0  \end{array} \right.
\]
where $\alpha$ is a primitive element in $\F_q$.
\end{enumerate}

\subsection{Multiplicative characters}

\begin{df}\label{df:mul_log}
Let $\alpha$ be a primitive element in $\F_{q}$ and $M$ a divisor of $q-1$, i.e., $M \mid q-1$.
We define a \emph{multiplicative character} of $\F_{q}$
of order $M$ by
\begin{equation}\label{eq:mul_log}
\psi(x) = \exp \left(j \frac{ 2 \pi \log_\alpha x}{M} \right), \quad x \in \F_{q}
\end{equation}
where $\psi(0) = 1$ by the definition of the log operation.
\end{df}
\vspace{0.075in}

For the original definition of multiplicative characters, see \cite{Lidl:FF}.
In (\ref{eq:mul_log}), note that $\psi(0) = 1$, which contradicts the conventional assumption in \cite{Lidl:FF}.
In this paper, however, we keep the assumption of $\psi(0) = 1$ to maintain the definition of (\ref{eq:mul_log}),
which is convenient for this work.
Throughout this paper, $\psi(x)$ may be denoted as $\psi$ if the context is clear.

The Weil bound~\cite{Weil:BNT} gives an upper bound on the magnitude of multiplicative character sums.
We introduce the refined version (Corollary 1 in~\cite{YuGong:ReWeil})
supporting the assumption $\psi(0) = 1$.

\vspace{0.075in}
\begin{prop}\label{prop:Weil_mod}\cite{YuGong:ReWeil}
Let $f_1(x),  \cdots, f_l (x)$ be $l$ monic and irreducible polynomials in $\F_q [x]$
which have positive degrees $d_1, \cdots, d_l$, respectively.
Let $d$ be the number of distinct
roots of $ f(x)= \prod_{i=1} ^lf_i(x)$ in its splitting field over $\F_q$.
Then, $d \leq \sum_{i=1} ^l d_i $, where the equality is achieved if $f_i(x)$'s are distinct.
Let $\psi_1, \cdots, \psi_l$ be multiplicative characters of $\F_q$.
Assume that the product character $ \prod_{i=1} ^l \psi_i (f_i(x))$ 
is nontrivial, i.e., $ \prod_{i=1} ^l \psi_i (f_i(x)) \neq 1$ for some $x \in \F_q$.
Let $e_i $ be the number of distinct roots in $\F_q$ of $f_i (x)$,
where $i = 1, \cdots, l$.
If $\psi_i(0) = 1$ for each $i$, then for every $a_i \in \F_q ^* , \ i=1, \cdots, l$, we have
\begin{equation}\label{eq:Weil_1_mod}
\left| \sum_{x\in \F_q} \psi_1 (a_1 f_1(x)) \cdots  \psi_l (a_l f_l(x) ) \right| \leq \left(d -1 \right) \sqrt{q}
+ \sum_{i=1} ^l e_i .
\end{equation}
\end{prop}
\vspace{0.075in}

In Proposition~\ref{prop:Weil_mod}, $d$ in (\ref{eq:Weil_1_mod}) 
can be replaced by $\sum_{i=1} ^l d_i $, since $d \leq \sum_{i=1} ^l d_i$.
The replacement allows us to have no need of distinguishing whether or not the polynomials are distinct,
which is useful for this work.

%
%


\subsection{Restricted isometry property}

The restricted isometry property (RIP)~\cite{CanTao:univ} presents a sufficient condition for a measurement matrix $\Abu$
to guarantee unique sparse recovery.
\vspace{0.075in}
\begin{df}\label{def:rip}
The restricted isometry constant $\delta_s$ of a $K \times N$ matrix $\Abu$ is defined
as the smallest number such that
\begin{equation*}\label{eq:rip}
(1-\delta_s) || \xbu ||_{l_2} ^2 \leq || \Abu \xbu ||_{l_2} ^2 \leq (1+\delta_s) || \xbu ||_{l_2} ^2
\end{equation*}
holds for all $s$-sparse vectors $\xbu \in \R^N$,
where $||\xbu ||_{l_2} ^2 = \sum_{n=0} ^{N-1} |x_n|^2$ with
$\xbu = (x_0, \cdots, x_{N-1})^T$.
\end{df}
\vspace{0.075in}
We say that $\Abu$ obeys the RIP of order $s$ if $\delta_s$ is reasonably small, not close to $1$.
In fact, the RIP requires that
all subsets of $s$ columns taken from the measurement matrix should be \emph{nearly orthogonal}~\cite{CanWak:intro}.
Indeed, Candes~\cite{Can:rip} asserted that if $\delta_{2s} <1$, a unique $s$-sparse solution
is guaranteed by $l_0$-minimization, which is however a hard combinatorial problem.

A tractable approach for sparse recovery is to solve the $l_1$-minimization, i.e.,
to find a solution of
$\min_{\widetilde{\xbu} \in \R^N} || \widetilde{\xbu} ||_{l_1}$ subject to $\Abu \widetilde{\xbu} = \ybu$,
where $||\widetilde{\xbu} ||_{l_1} = \sum_{i=0} ^{N-1} |\widetilde{x}_i|$.
In addition,
greedy algorithms~\cite{Tropp:greed} have been also proposed
for sparse signal recovery,
including matching pursuit (MP)~\cite{Mallat:mp}, orthogonal matching pursuit (OMP)~\cite{Tropp:omp},
and CoSaMP~\cite{Needell:cosamp}.
If a deterministic measurement matrix is used, 
its structure may be exploited to develop
a reconstruction algorithm for sparse signal recovery,
providing fast processing and low complexity~\cite{App:chirp}\cite{How:fast}.

\subsection{Coherence and redundancy}
In compressed sensing, a $K \times N$ deterministic matrix $\Abu$ is associated with
two geometric quantities,
\emph{coherence}
and \emph{redundancy}~\cite{Tropp:gap}.
The coherence $\mu$ is defined by
\begin{equation*}
\mu = \max_{0 \leq l \neq m \leq N-1} \left| \abu_l ^H \cdot \abu_m \right|
\end{equation*}
where $\abu_*$ denotes a column vector of $\Abu$ with $|| \abu_* ||_{l_2} = 1$,
and $\abu_* ^H$ is its conjugate transpose.
In fact, the coherence is a measure of mutual orthogonality among columns,
and the small coherence is desired for good sparse recovery~\cite{Rauhut:struct}.
In general, the coherence is lower bounded by
\begin{equation*}\label{eq:welch}
\mu \geq \sqrt{\frac{N-K}{K(N-1)}}
\end{equation*}
which is called the \emph{Welch bound}~\cite{Welch:low}.

The redundancy, on the other hand, is defined as $\rho = || \Abu ||^2$,
where $|| \cdot ||$ denotes the spectral norm of $\Abu$,
or the largest singular value of $\Abu$.
We have $\rho \geq N/K$, where the equality holds if and only if $\Abu$
is a \emph{tight} frame~\cite{Cald:LASSO}.
For unique sparse recovery, it is also desired
that $\Abu$ should be a tight frame with the smallest redundancy~\cite{Cald:LASSO}.

\section{Compressed Sensing Matrices from Multiplicative Character Sequences}

Sidelnikov~\cite{Sidel:org} introduced two types of polyphase sequences with low periodic autocorrelation.
In what follows, we define the sequences by logarithm and construct compressed sensing matrices
using the sequences.

\subsection{Construction from power residue sequences}

\begin{df}\label{df:prs}
Let $p$ be an odd prime and $M$ a divisor of $p-1$, i.e., $M \mid p-1$.
Let $\alpha$ be a primitive root modulo $p$.
An $M$-ary power residue sequence $\rbu = \{ r(k) \mid 0 \leq k \leq p-1 \}$
of period $p$ is defined by
\begin{equation}\label{eq:prs_log}
r(k) \equiv  \log_\alpha k \mod M . 
\end{equation}
By (\ref{eq:mul_log}) and (\ref{eq:prs_log}), the modulated sequence of $r(k)$ is represented by
\begin{equation*}\label{eq:prs_mod}
\omega_M ^{r(k)} = \psi(k), \quad 0 \leq k \leq p-1
\end{equation*}
where $\psi(0)=1$.
\end{df}
\vspace{0.075in}

Employing the power residue sequence, we construct a compressed sensing matrix.

\vspace{0.075in}
\begin{const}\label{cst:mat_prs}
Let $\rbu = \{ r(k) \mid 0 \leq k \leq p-1 \}$ be an $M$-ary power residue sequence of period $p$, where $M>2$.
Let $K = p$ and $N = (M-1)K$.
In a $K \times N$ matrix $\Abu$,
set each column index as $n = (c-1)K+b$, where
$b \equiv n \mod{K}$ and $ c = \left\lfloor \frac{n}{K} \right\rfloor + 1$ for
$0\leq b \leq p-1$ and $1 \leq c \leq M-1$.
Then, we construct
a $K \times N$ compressed sensing matrix $\Abu$ where each entry is given by
\begin{equation*}\label{eq:phi}
\begin{split}
a_{k, n}  = \frac{1}{\sqrt{K}} \omega_M ^{c r(k+b)} = \frac{1}{\sqrt{K}} \psi \left( (k+b)^c \right) & , \\
0 \leq k \leq K-1, \ 0 \leq n \leq N-1 &
\end{split}
\end{equation*}
where
$k+b$ is computed modulo $p$.
\end{const}
\vspace{0.075in}


\begin{thr}\label{th:coh_mul}
For the $K \times N$ matrix $\Abu$ in Construction~\ref{cst:mat_prs},
the coherence is given by
\begin{equation}\label{eq:coh_mul}
\mu = \max_{0 \leq n_1 \neq n_2 \leq N-1} \left| \abu_{n_1} ^H \cdot \abu_{n_2} \right| =
\frac{\sqrt{K}+2}{K}
\end{equation}
where $\abu_*$ denotes a column vector of $\Abu$.
In particular, if $K$ and $M$ are large,
the coherence is
asymptotically optimal achieving the equality of the Welch bound.
\end{thr}
\vspace{0.075in}

\noindent \textit{Proof.}
Consider the column indices of $n_1 = (c_1-1)K+b_1$ and $n_2 = (c_2-1)K+b_2$, where $n_1 \neq n_2$.
Set $x=k \in \F_p$. Then,
the Weil bound in Proposition~\ref{prop:Weil_mod} gives an upper bound on
the magnitude of the inner product of a pair of columns in $\Abu$, i.e.,
\begin{equation*}
\begin{split}
\left| \abu_{n_1} ^H \cdot \abu_{n_2} \right|
& = \frac{1}{K} \left| \sum_{x \in \F_p} \psi\left( (x+b_1)^{-c_1} \right) \cdot \psi\left( (x+b_2)^{c_2} \right) \right| \\
& = \frac{1}{K} \left| \sum_{x \in \F_p} \psi_1 \left( x+b_1 \right) \psi_2 \left( x+b_2 \right) \right| \\
& \leq \frac{\sqrt{K}+2}{K} 
\end{split}
\end{equation*}
where $\psi_1 = \psi^{-c_1}$ and $\psi_2 = \psi^{c_2}$.
Therefore, the coherence in (\ref{eq:coh_mul}) is immediate. 
For large $K$, $\mu \approx \frac{1}{\sqrt{K}}$ from (\ref{eq:coh_mul}).
Also, the equality of the Welch bound is 
$ \sqrt{\frac{N-K}{K(N-1)}} = \sqrt{\frac{M-2}{(M-1)K-1}} \approx \frac{1}{\sqrt{K}}$ for large $M$.
Thus, the coherence
asymptotically achieves the equality of the Welch bound for large $K$ and $M$.
\qed
\vspace{0.075in}

Tropp~\cite{Tropp:greed} 
revealed that $(2s-1) \mu<1$ ensures $s$-sparse signal recovery by
basis pursuit (BP) and orthogonal matching pursuit (OMP).
With the sufficient condition, 
Gribonval and Vandergheynst~\cite{Grib:MP} further showed that matching pursuit (MP)
also derives a unique solution with exponential convergence.
Using the results, Theorem~\ref{th:s_cds} provides a sufficient condition for the matrix $\Abu$ in Construction~\ref{cst:mat_prs}.
\vspace{0.075in}
\begin{thr}\label{th:s_cds}
For the matrix $\Abu$ in Construction~\ref{cst:mat_prs},
a unique $s$-sparse solution is guaranteed by $l_1$-minimization or greedy algorithms if
\begin{equation*}\label{eq:s_cds}
s< \frac{1}{2} \left(\frac{K}{\sqrt{K}+2} + 1 \right).
\end{equation*}
\end{thr}
\vspace{0.075in}
\noindent \textit{Proof.}
The upper bound on the sparsity level 
is straightforward from the coherence $\mu = \frac{\sqrt{K}+2}{K}$
and the Tropp's condition $(2s-1) \mu<1$.
\qed

\subsection{Construction from Sidelnikov sequences}

\begin{df}\label{df:Sidel}
Let $\alpha$ be a primitive element in the finite field $\F_{p^m}$  
and $M$ a divisor of $p^m-1$, 
where $p$ is prime and $m$ is a positive integer.
An $M$-ary Sidelnikov sequence $\sbu = \{ s(k) \mid 0 \leq k \leq p^m-2 \}$
of period $p^m-1$ is defined by
\begin{equation}\label{eq:Sidel_log}
s(k) \equiv \log_\alpha (\alpha^k+1) \mod M . 
\end{equation}
By (\ref{eq:mul_log}) and (\ref{eq:Sidel_log}), the modulated sequence of $s(k)$ is represented by
\begin{equation*}\label{eq:Sidel_mod}
\omega_M ^{s(k)} = \psi(\alpha^k+1), \quad 0 \leq k \leq p^m-2
\end{equation*}
where $\psi(0)=1$.
\end{df}
\vspace{0.075in}


\begin{table*}[t!]
\caption{The spectral norms of $\Abu$ and $\widehat{\Abu}$}
\fontsize{9}{14pt}\selectfont
\label{tb:spect}
\centering
\begin{tabular}{|c|c|c||c|c|c|c|}
\hline
$(K, N, M)$    &   $||\Abu||$  &  $\sqrt{N/K}$ &     $(K, N, M)$   &   $||\widehat{\Abu}||$  &  $\sqrt{N/K}$ \\
\hline
\hline
$(43, 1763, 42)$ &  $6.6282$   & $6.4031$                &     $(26, 650, 26)$  & $5.0990$             &  $5$   \\
$(59, 3363, 58)$ &  $7.7431$   &  $7.5498$               &     $(48, 2256, 48)$ & $6.9282$             &  $6.8557$  \\
$(67, 4355, 66)$ &  $8.2439$   &  $8.0623$               &     $(80, 6320, 80)$  & $8.9443$            &  $8.8882$  \\
$(83, 6723, 82)$ &  $9.1635$   &  $9$                    &     $(124, 15252, 124)$ & $11.1355$         &  $11.0905$  \\
$(97, 9215, 96)$ &  $9.8982$   &  $9.7468$               &     $(168, 28056, 168)$ & $12.9615$         &  $12.9228$  \\
\hline
\end{tabular}
\end{table*}

\begin{const}\label{cst:mat_Sidel}
Let $\sbu = \{ s(k) \mid 0 \leq k \leq p^m-2 \}$ be an $M$-ary Sidelnikov sequence of period $p^m-1$, where $M>2$.
Let $K = p^m-1$ and $N = (M-1)K$.
In a $K \times N$ matrix $\widehat{\Abu}$,
set each column index as $n = (c-1)K+b$, where
$ b \equiv n \mod{K}$ and $ c = \left\lfloor \frac{n}{K} \right\rfloor + 1$
for $0\leq b \leq p^m-2$ and $1 \leq c \leq M-1$.
Then, we construct a $K \times N$ compressed sensing matrix $\widehat{\Abu}$ where each entry is given by
\begin{equation*}\label{eq:phi_s}
\begin{split}
\widehat{a}_{k, n} = \frac{1}{\sqrt{K}} \omega_M ^{c s(k+b)} = \frac{1}{\sqrt{K}} \psi \left( (\lambda \alpha^k+1)^c \right) & , \\
0 \leq k \leq K-1, \ 0 \leq n \leq N-1 &
\end{split}
\end{equation*}
where $\lambda = \alpha^b \in \F_{p^m} ^*$ and
$k+b$ is computed modulo $p^m-1$.
\end{const}

\vspace{0.075in}
\begin{thr}\label{th:coh_Sidel}
For the $K \times N$ matrix $\widehat{\Abu}$ in Construction~\ref{cst:mat_Sidel},
the coherence is given by
\begin{equation}\label{eq:coh_Sidel}
\mu = \max_{0 \leq n_1 \neq n_2 \leq N-1} \left| \widehat{\abu}_{n_1} ^H \cdot \widehat{\abu}_{n_2} \right| =
\frac{\sqrt{K+1}+3}{K}.
\end{equation}
where $\widehat{\abu}_*$ denotes a column vector of $\widehat{\Abu}$.
For large $K$ and $M$,
the coherence is
asymptotically optimal achieving the equality of the Welch bound.
\end{thr}
\vspace{0.075in}
\noindent \textit{Proof.}
Consider the column indices of $n_1 = (c_1-1)K+b_1$ and $n_2 = (c_2-1)K+b_2$, where $n_1 \neq n_2$.
Let $\lambda_1 = \alpha^{b_1}$, $\lambda_2 = \alpha^{b_2}$, and
$x = \alpha^k \in \F_{p^m} ^*$.
Using the Weil bound in Proposition~\ref{prop:Weil_mod}, the inner product of a pair of columns in $\Abu$ has the magnitude bounded by
\begin{equation*}
\begin{split}
\left| \widehat{\abu}_{n_1} ^H \cdot \widehat{\abu}_{n_2} \right|
& = \frac{1}{K} \left| \sum_{x \in \F_{p^m} ^* } \psi\left( (\lambda_1x+1)^{-c_1} \right) \cdot \psi\left( (\lambda_2x+1)^{c_2} \right) \right| \\
& = \frac{1}{K} \left| \sum_{x \in \F_{p^m} ^*} \psi_1 \left( \lambda_1x+1 \right) \psi_2 \left( \lambda_2x+1 \right) \right| \\
& = \frac{1}{K} \left( \left| \sum_{x \in \F_{p^m} } \psi_1 \left( \lambda_1x+1 \right) \psi_2 \left( \lambda_2x+1 \right) \right|  + 1 \right) \\
& \leq \frac{\sqrt{K+1}+3}{K} 
\end{split}
\end{equation*}
where $\psi_1 = \psi^{-c_1}$ and $\psi_2 = \psi^{c_2}$.
Therefore, the coherence in (\ref{eq:coh_Sidel}) is immediate. 
Similar to the approach made in the proof of Theorem~\ref{th:coh_mul},
the coherence asymptotically achieves
the equality of the Welch bound for large $K$ and $M$.
\qed
\vspace{0.075in}

With the coherence and the Tropp's sufficient condition~\cite{Tropp:greed},
the sparsity bound of $\widehat{\Abu}$
is straightforward.

\vspace{0.075in}
\begin{thr}\label{th:s_Sidel}
For the matrix $\widehat{\Abu}$ in Construction~\ref{cst:mat_Sidel},
a unique $s$-sparse solution is guaranteed by $l_1$-minimization or greedy algorithms if
\begin{equation*}\label{eq:s_Sidel}
s< \frac{1}{2} \left(\frac{K}{\sqrt{K+1}+3} + 1 \right).
\end{equation*}
\end{thr}

\vspace{0.075in}
\begin{rem}
In Constructions~\ref{cst:mat_prs} and \ref{cst:mat_Sidel},
it is easily checked that a pair of rows in the matrix $\Abu$ (or $\widehat{\Abu}$)
may not be mutually orthogonal, which implies that $\Abu$ (or $\widehat{\Abu}$) is not a tight frame.
However, numerical data reveals that their spectral norms are nearly optimal, 
as shown in Table~\ref{tb:spect} for some values of $K, N$, and $M$.
\end{rem}
\vspace{0.075in}

\begin{rem}
In Constructions~\ref{cst:mat_prs} and \ref{cst:mat_Sidel},
each column vector of $\Abu$ (or $\widehat{\Abu}$) is equivalent to a modulated sequence
of a constant multiple of a cyclic shift of a power residue
or Sidelnikov sequence.
Therefore, only a single base sequence is required
for the implementation of the compressed sensing matrix,
which allows low complexity and storage.
In particular, if $m$ is even, we can generate a Sidelnikov sequence of period $p^{m}-1$
using an efficient linear feedback shift register (LFSR)~\cite{YuGong:struct},
which further reduces the implementation complexity.
Moreover, the alphabet size $M$ of the matrix is variable depending on the signal dimension $N$,
as $N$ is a function of $M$ in the constructions.
In summary, we can efficiently implement a variety of compressed sensing matrices
with various alphabet and column sizes from multiplicative character sequences.
\end{rem}


\subsection{RIP analysis}

For statistical RIP analysis, we chose $(K, N, M) = (47, 2115, 46)$ for power residue sensing matrix $\Abu$,
and $(K, N, M) = (48, 2256, 48)$ for
Sidelnikov sensing matrix $\widehat{\Abu}$, respectively.
For each matrix, taking its submatrix with randomly chosen $s$ columns,
we then measured the condition number, defined as
the ratio of the largest singular value of each submatrix to the smallest.

\begin{figure}[!t]
\centering
\includegraphics[width=0.51\textwidth, angle=0]{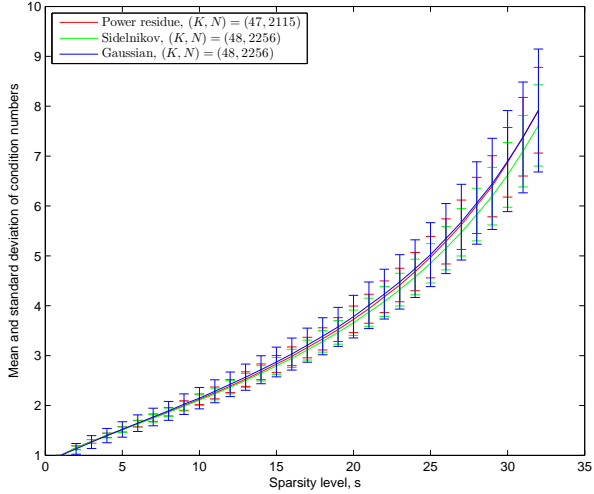}
\caption{Condition number statistics of the Gram matrices of power residue, Sidelnikov, and Gaussian random sensing matrices.}
\label{fig:eigen}
\end{figure}

In Figure~\ref{fig:eigen}, we measured the means and standard deviations of the condition numbers
of $\Abu_s$, $\widehat{\Abu}_s$, and $\Gbu_s $, where %
$\Abu_s$ and $\widehat{\Abu}_s$ are the submatrices of $s$ columns randomly chosen from $\Abu$ and $\widehat{\Abu}$, respectively, 
while $\Gbu_s$ is a $K \times s$ Gaussian random matrix with $K=48$.
The statistics were measured over total $10, 000$ condition numbers,
where each matrix is newly chosen at each instance.
Each entry of the Gaussian matrix $\Gbu_s$ is independently sampled from the Gaussian distribution
of zero mean and variance $\frac{1}{K}$, and each column vector is then normalized such that
it has unit $l_2$-norm.

Note that the singular values of $\Abu_s$ (or $\widehat{\Abu}_s$) are the square roots of eigenvalues of the Gram matrix $\Abu_s ^H \Abu_s$
(or $\widehat{\Abu}_s ^H \widehat{\Abu}_s$).
Then, the condition numbers should be as small as possible for sparse signal recovery,
since the RIP requires that the Gram matrix should have all the eigenvalues
in an interval $[1-\delta_s, 1+\delta_s]$ with reasonably small $\delta_s$~\cite{Rauhut:struct}.
From this point of view,
we see that
the power residue and Sidelnikov sensing matrices
show better statistics of condition numbers than the Gaussian random matrix in Figure~\ref{fig:eigen}.
We made similar observations in the statistics of power residue sensing matrix with $(K, N, M) = (127, 15875, 126)$
and Sidelnikov matrix with $(K, N, M) = (124, 15252, 124)$.
This convinces us that $\Abu$ and $\widehat{\Abu}$ in Constructions~\ref{cst:mat_prs} and \ref{cst:mat_Sidel}
are suitable for compressed sensing
in a statistical sense.

\section{Recovery performance}

\subsection{Recovery from noiseless data}

\begin{figure}[t!]
\centering
\includegraphics[width=0.51\textwidth, angle=0]{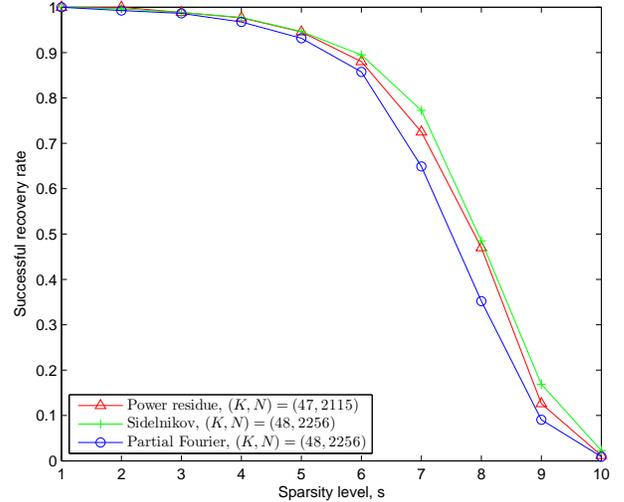}
\caption{Successful recovery rates for power residue, Sidelnikov, and partial Fourier sensing matrices
for noiseless sparse signals.}
\label{fig:succ_ps}
\end{figure}

Figure~\ref{fig:succ_ps} shows numerical results on successful recovery rates of $s$-sparse signals
measured by power residue sensing matrix $\Abu$ with $(K, N, M) = (47, 2115, 46)$,
and Sidelnikov matrix $\widehat{\Abu}$ with $(K, N, M) = (48, 2256, 48)$, respectively,
where total $2000$ sample vectors were tested for each sparsity level.
For comparison, the figure also displays the rate for $48 \times 2256$ randomly chosen partial Fourier matrix,
where a new matrix is used at each instance
of an $s$-sparse signal,
in order to obtain the average rate.
For all the matrices, the matching pursuit recovery with $100$ iterations has been applied for the reconstruction of sparse signals.
Each nonzero entry of an $s$-sparse signal $\xbu$ has the magnitude of $1$,
where its position and sign are chosen uniformly at random.
A success is declared in the reconstruction if the squared error is reasonably small for the estimate $\widehat{\xbu}$, i.e.,
$|| \xbu - \widehat{\xbu} ||_{l_2} ^2 < 10^{-4}$.

In the figure, we observe that if $s\leq 3$, more than $99 \%$ of $s$-sparse signals
are successfully recovered for the power residue sensing matrix,
which statistically verifies the sufficient condition in Theorem~\ref{th:s_cds}.
Similarly, the sufficient condition of the Sidelnikov sensing matrix in Theorem~\ref{th:s_Sidel} is also verified
for $s \leq 2$.
In fact, the figure reveals that the sufficient conditions are somewhat pessimistic,
since the actual recovery performance is fairly good even for high sparsity levels.
For example, both sensing matrices guarantee more than $95 \%$ successful recovery rates if $s \leq 4$.
Furthermore, the matrices present better recovery performance than randomly chosen partial Fourier matrices. 
We made similar observations in the recovery performance of other power residue ($(K, N, M) = (127, 15875, 126)$)
and Sidelnikov ($(K, N, M) = (124, 15252, 124)$) sensing matrices.

\subsection{Recovery from noisy data}

\begin{figure}[t!]
\centering
\includegraphics[width=0.51\textwidth, angle=0]{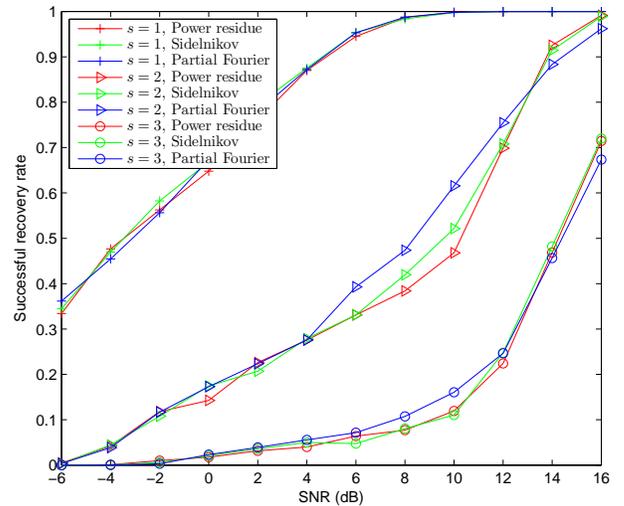}
\caption{Successful recovery rates for power residue, Sidelnikov, and partial Fourier sensing matrices in the presence of noise.
($(K, N)=(47, 2115)$ for power residue and $(K, N)=(48, 2256)$ for the others)}
\label{fig:succ_n}
\end{figure}

In practice, a measured signal $\ybu$ contains noise, i.e.,
$\ybu = \Abu \xbu + \zbu$, where $\zbu \in \C^K$ denotes a $K$-dimensional complex vector of noise.
Thus, a compressed sensing matrix must be robust to the measurement noise
for stable and noise resilient recovery.
Figure~\ref{fig:succ_n} displays the matching pursuit recovery performance of
various compressed sensing matrices in the presence of noise.
The matrix parameters and the sparse signal generation are identical to those of noiseless case.
In the figure, $\xbu$ is $s$-spare for $s=1, 2,3$, and
the signal-to-noise ratio (SNR) is defined by
${\rm SNR} 
= \frac{||\Abu \xbu ||_{l_2} ^2}{K \sigma_z^2}$,
where each element of $\zbu$ is an independent and identically distributed (i.i.d.)
complex Gaussian random process with zero mean and variance $\sigma_z ^2$.
In noisy recovery, a success is declared if 
$|| \xbu - \widehat{\xbu} ||_{l_2} ^2 < 10^{-2}$ after $100$ iterations.
From Figure~\ref{fig:succ_n}, we observe that the recovery performance 
is stable and robust against noise corruption at sufficiently high SNR,
which is
similar to that of randomly chosen partial Fourier matrices.

\section{Conclusion}
This paper has presented how to deterministically construct a $K \times N$ measurement matrix for compressed sensing
via multiplicative character sequences.
We showed that the matrices from $M$-ary power residue and Sidelnikov sequences
have asymptotically optimal coherence for large $K$ and $M$.
We also presented the sufficient conditions
on the sparsity level for unique sparse solution.
Furthermore, the RIP of the matrices has been statistically analyzed,  
where we observed that they have better condition number statistics than Gaussian random matrices.
Numerical results revealed that
the compressed sensing matrices 
show stable and reliable performance in matching pursuit recovery
for sparse signals with or without measurement noise.
Finally, we would like to mention that the compressed sensing matrices can be implemented with small storage and low complexity,
as each column vector is equivalently generated by a constant multiple of a cyclic shift of a single base power residue
or Sidelinikov sequence.







\begin{thebibliography}{1}







\bibitem{Donoho:CS}
D.~L.~Donoho,
``Compressed Sensing,''
\emph{IEEE Trans. Inf. Theory}, vol.~52, no.~4, pp.~1289-1306, Apr. 2006.


\bibitem{CanRomTao:robust}
E.~J.~Candes, J.~Romberg, and T.~Tao,
``Robust uncertainty principles: exact signal reconstruction from highly incomplete frequency information,''
\emph{IEEE Trans. Inf. Theory}, vol.~52, no.~2, pp.~489-509, Feb. 2006.



\bibitem{CanTao:univ}
E.~J.~Candes and T.~Tao,
``Near-optimal signal recovery from random projections: universal encoding strategies,''
\emph{IEEE Trans. Inf. Theory}, vol.~52, no.~12, pp.~5406-5425, Dec. 2006.



\bibitem{Rauhut:struct}
H.~Rauhut,
``Compressive sensing and structured random matrices,''
\emph{preprint.} May 2010.




\bibitem{Cald:large}
R.~Calderbank, S.~Howard, and S.~Jafarpour,
``Construction of a large class of deterministic sensing matrices that satisfy a statistical isometry property,''
\emph{IEEE Journal of Selected Topics in Signal Processing}, vol.~4, no.~2, pp.~358-374. Apr. 2010.




\bibitem{App:chirp}
L.~Applebaum, S.~D.~Howard, S.~Searle, and R.~Calderbank,
``Chirp sensing codes: deterministic compressed sensing measurements for fast recovery,''
\emph{Appl. and Comput. Harmon. Anal.}, vol.~26, pp.~283-290, 2009.

\bibitem{Cald:sub}
R.~Calderbank, S.~Howard, and S.~Jafarpour,
``A sublinear algorithm for sparse reconstruction with $l_2/l_2$ recovery guarantees,''
\emph{arXiv:0806.3799v2 [cs.IT]}, Oct. 2009.


\bibitem{How:fast}
S.~Howard, R.~Calderbank, and S.~Searle,
``A fast reconstruction algorithm for deterministic compressive sensing using
second order Reed-Muller codes,''
\emph{Conference on Information Systems and Sciences (CISS)}, Princeton, NJ, Mar. 2008.


\bibitem{Ailon:BCH}
N.~Ailon and E.~Liberty,
``Fast dimension reduction using Rademacher series on dual BCH codes,''
\emph{Annual ACM-SIAM Symposium on Discrete Algorithms (SODA)}, pp.~215-224, Jan.~2008.

\bibitem{DeVore:det}
R.~A.~DeVore,
``Deterministic constructions of compressed sensing matrices,''
\emph{Journal of Complexity}, vol.~28, pp.~918-925, 2007.

\bibitem{Gur:osc}
S.~Gurevich, R.~Hadani, and N.~Sochen,
``On some determistic dictionaries supporting sparsity,''
\emph{Journal of Fourier Analysis and Applications}, To appear.


\bibitem{Xu:trig}
Z.~Xu,
``Deterministic sampling of sparse trigonometric polynomials,''
\emph{arXiv:1006.2221v1 [math.NA]}, June 2010.



\bibitem{Yu:add}
N.~Y.~Yu,
``Deterministic compressed sensing matrices from additive character sequences,''
\emph{arXiv:1010.0011v1 [cs.IT]}, Oct. 2010.


\bibitem{Yu:CDS}
N.~Y.~Yu,
``Deterministic construction of partial Fourier compressed sensing matrices via cyclic difference sets,''
\emph{arXiv:1008.0885v1 [cs.IT]}, Aug. 2010.



\bibitem{Lidl:FF}
R.~Lidl and H.~Niederreiter,
\emph{Finite Fields,} in
\emph{Encyclopedia of Mathematics and Its Applications},
vol.~20, Cambridge University Press, 1997.


\bibitem{Weil:BNT}
A.~Weil,
\emph{Basic Number Theory}, 3rd. Ed., Springer-Verlag, 1974.





\bibitem{YuGong:ReWeil}
N.~Y.~Yu and G.~Gong,
``Multiplicative characters, the Weil bound, and
polyphase sequence families with low correlation,''
\emph{IEEE Trans. Inf. Theory}, To appear.
Also available at CACR 2009-25, \emph{CACR Technical Report}, University of Waterloo, 2009.























\bibitem{CanWak:intro}
E.~J.~Candes and M.~B.~Wakin,
``An introduction to compressive sampling,''
\emph{IEEE Sig. Proc. Mag.}, pp.~21-30, Mar. 2008.




\bibitem{Can:rip}
E.~J.~Candes,
``The restricted isometry property and its implications for compressed sensing,''
\emph{Academie des sciences}, Feb. 2008.


\bibitem{Tropp:greed}
J.~A.~Tropp,
``Greed is good: algorithmic results for sparse approximation,''
\emph{IEEE Trans. Inf. Theory}, vol.~50, no.~10, pp.~2231-2242, Oct. 2004.




\bibitem{Mallat:mp}
S.~Mallat and Z.~Zhang,
``Matching pursuits with time-frequency dictionaries,''
\emph{IEEE Trans. Singal Processing}, vol.~41, pp.~3397-3415, Dec. 1993.

\bibitem{Tropp:omp}
J.~A.~Tropp and A.~C.~Gilbert,
``Signal recovery from random measurements via orthogonal matching pursuit,''
\emph{IEEE Trans. Inf. Theory}, vol.~53, no.~12, pp.~4655-4666, Dec. 2007.


\bibitem{Needell:cosamp}
D.~Needell and J.~A.~Tropp,
``CoSaMP: Iterative signal recovery from incomplete and inaccurate samples,''
\emph{Appl. and Comput. Harmon. Anal.}, vol.~26, pp.~301-321, 2009.


\bibitem{Tropp:gap}
J.~A.~Tropp,
``The sparsity gap: uncertainty principle proportional to dimension,''
\emph{Conference on Information Systems and Sciences (CISS)}, Princeton, NJ, Mar. 2010.





\bibitem{Welch:low}
L.~R.~Welch,
``Lower bounds on the maximum cross correlation of signals,''
\emph{IEEE Trans. Inf. Theory}, vol.~IT-20, no.~3, pp.~397-399, May 1974.




\bibitem{Cald:LASSO}
R.~Calderbank and S.~Jafarpour,
``Reed Muller sensing matrices and the LASSO,''
\emph{arXiv:1004.4949v1 [cs.IT]}, Apr. 2010.


\bibitem{Sidel:org}
V.~M.~Sidelnikov,
``Some $k$-valued pseudo-random sequences and nearly equidistant codes,''
\emph{Probl. Inf. Transm.}, vol.~5, pp.~12-16, 1969.



\bibitem{Grib:MP}
R.~Gribonval and P.~Vandergheynst,
``On the exponential convergence of matching pursuits in quasi-incoherent dictionaries,''
\emph{IEEE Trans. Inf. Theory}, vol.~52, no.~1, pp.~255-261, Jan. 2006.



\bibitem{YuGong:struct}
N.~Y.~Yu and G.~Gong,
``New construction of M-ary sequence families with low correlation from the structure of Sidelnikov sequences,''
\emph{IEEE Trans. on Inf. Theory}, vol.~56, no.~8, pp. 4061-4070, Aug. 2010.


































\end{thebibliography}
%

\end{document}